\documentclass[12pt]{iopart}
\usepackage{graphicx}
\usepackage{amsmath}
\newcommand{\erf}{\mathop{\mathrm{erf}}}
\begin{document}

\title[]{Modeling Heavy-Ion Fusion Cross Section Data via a Novel Artificial Intelligence Approach}

\author{Daniele Dell'Aquila$^{2,3,*}$, Brunilde Gnoffo$^1$, Ivano Lombardo$^{1,\dagger}$, Francesco Porto$^{3,4}$ and Marco Russo$^{1,4}$}
\address{1 INFN - Sezione di Catania, Catania, Italy}
\address{2 Dipartimento di Chimica e Farmacia, Universit\`{a} degli Studi di Sassari, Sassari, Italy}
\address{3 INFN - Laboratori Nazionali del Sud, Catania, Italy}
\address{4 Dipartimento di Fisica e Astronomia, Universit\`{a} degli Studi di Catania, Catania, Italy}

\ead{$^*$ddellaquila@uniss.it; $^\dagger$ivano.lombardo@ct.infn.it}
\vspace{10pt}
\begin{indented}
\item[] Submitted: January 2022
\end{indented}

\begin{abstract}
We perform a comprehensive analysis of complete fusion cross section data with the aim to derive, in a completely data-driven way, a model suitable to predict the integrated cross section of the fusion between light to medium mass nuclei at above barrier energies. To this end, we adopted a novel artificial intelligence approach, based on a hybridization of genetic programming and artificial neural networks, capable to derive an analytical model for the description of experimental data. The approach enables, for the first time, to perform a global search for computationally simple models over several variables and a considerable body of nuclear data. The derived phenomenological formula can serve to reproduce the trend of fusion cross section for a large variety of light to intermediate mass collision systems in an energy domain ranging approximately from the Coulomb barrier to the onset of multi-fragmentation phenomena.
\end{abstract}

\vspace{2pc}

\section{Introduction}
\label{sec:introduction}
The study of heavy-ion fusion at energies well above the Coulomb barrier has been the subject of extended investigations, especially during the $'70$s and the $'90$s of the past century \cite{Glas75,Bass80,Sanders99,Esbensen14,Jha20}. The use of different experimental methods to estimate the yields of emitted evaporation residues (gamma-ray analysis, time-of-flight and magnetic spectrometers, charged particle detection with telescope arrays) made it possible to collect a large body of data, in particular at energies ranging from the Coulomb barrier up to the dominance of incomplete fusion and the onset of multi-fragmentation \cite{Frobrich84}. The analysis of these data has been performed mainly in terms of semi-classical models: for light-to-medium systems (i.e., for compound nuclei in the range $A_{tot}\simeq20-140$, where the presence of fusion-fission phenomena can be, in first approximation, neglected) the fusion cross section, usually plotted as a function of the inverse of the center-of-mass energy $1/E_{cm}$, can be very schematically divided into three regions (namely I, II, III) \cite{Bass80,Lee80}. The region I starts approximately at energies corresponding to the Coulomb barrier; in this region, the fusion cross section increases almost linearly for decreasing values of $1/E_{cm}$, until reaching a smooth \emph{plateau} around a maximum value or a sudden change of slope. In region I, the fusion cross section exhausts almost all the flux of the reaction cross section \cite{Glas75,Lee80,Bass80,Pakou15,Mazzocco15}. The region II, starting from the end of region I, is typically characterized by a smooth fall of the fusion cross section for decreasing values of $1/E_{cm}$. It is often difficult to find the boundaries of region II: in fact, at large $E_{cm}$ values, the measurements of fusion cross sections became more and more sophisticated, and the reported data are often affected by large uncertainties. After the smooth fall of region II, the fusion cross section still decreases for decreasing values of $1/E_{cm}$, in an energy regime where the fusion-evaporation mechanism gives rise to incomplete fusion mechanisms and, in general, to much more complicated reaction scenarios \cite{Jha20}. This part of the excitation function is often referred as the region III. At very large $E_{cm}$ values, the fusion cross section tends to zero, due to mechanical and thermodynamical instabilities arising in the transient system formed in heavy-ion central collisions \cite{Amorini09,Cardella12,DeFilippo09,Borderie18,Manduci16,Eudes14,Giordano90}.

For the sake of concision, we will indicate in the rest of the manuscript as the “region 0” the energy region in which fusion cross section is fully dominated by the effect of tunneling through the Coulomb barrier (\textit{sub-barrier} fusion). This part of data has been the subject of a great interest in the last two decades \cite{Esbensen14} and many effects of nuclear structure on fusion reaction mechanism have been enlightened through the analysis of such data. 

On the contrary, after two decades of intense experimental work ($’70$s-$’90$s), the studies of heavy-ion fusion well above the barrier (i.e., in the regions II and III) were largely neglected, leaving unsolved several aspects that could be explained by using modern detection systems now available \cite{DellAquila18,Acosta16,Pastore17,DellAquila19}. For example, the existence of a limiting angular momentum in fusion was explicitly investigated only for a few systems \cite{Beck96}. Such a phenomenon is linked with the data of regions II and III. To interpret this portion of excitation functions, two main families of macroscopic models have been reported: critical distance models, based on the occurrence of entrance channel effects \cite{Lee80,Matsuse82} and models based on limitations to compound nucleus \cite{bass77}.
 
Beyond such models, also microscopical approaches (as TDHF, see e.g. \cite{Reinhard16,Zheng18}), molecular dynamics (see e.g. \cite{Maruyama02,Feng08}), and phenomenological models (see, e.g., \cite{Glas75,Horn78,Lozano80,Kailas81,Giordano90,Eudes14}) have been developed, the last ones with the aim to describe large datasets of heavy ion fusion in broad mass and energy domains. Phenomenological models are often based on starting hypotheses inspired by nuclear reaction theory and are then analytically adapted to derive simple formulas able to describe data in the broadest possible domain. As an example, in Refs. \cite{Porto84,Giordano90,Horn78,Kailas81,Lozano80} are reproduced in a satisfactory way complete fusion data from several collision systems in the I-II regions, while in Ref. \cite{Eudes14} the description of data is limited to region III, where complete fusion is in competition with different reaction mechanisms \cite{Pirr01}. Such approaches are useful to make numerical evaluations on fusion cross sections but have some limitations: they often describe just a limited part of the energy regimes where fusion can occur, or they are lacking physical boundaries. For example, a phenomenological approach to fusion should guarantee that the fusion cross section goes toward zero at very low bombarding energies (deep sub-barrier regime) and at very high energies (where incomplete fusion and vaporization would be dominant in central reaction events), but the continuous evolution of the reaction mechanisms as a function of energy and the impact of quantum effects (such as tunnelling through barriers at low energies) make it difficult to fulfill these requirements (see, e.g, \cite{Porto84,Giordano82,Giordano90}).

Other interesting questions that can arise from the analysis of fusion data are linked with the possible effects of nuclear structure parameters (i.e. characteristics of projectiles, targets and compound nuclei) on fusion cross sections. These effects can be described introducing additional variables in the models. However, the complexity of the deriving datasets (see, e.g. \cite{Frobrich84}) alongside with the large variability of measured cross sections values (typically several orders of magnitudes) make it extremely challenging to consistently describe the complete fusion cross section by using ordinary fitting procedures.

Considering all these aspects, we decided to use a novel artificial intelligence approach to analyze a comprehensive dataset of light to medium mass fusion data available in the literature. Artificial intelligence has been already proven to be particularly informative to investigate nuclear physics variables and the outcome of nuclear experiments (see. e.g. Refs.~\cite{Neudecker21,Bai21}). The technique used in this paper, which is based on a state-of-the-art hybridization of genetic programming and artificial neural networks \cite{Russo16,Russo20}, has been already successfully applied to several research fields, including power plant engineering \cite{RussoSolar}, audio analysis \cite{Campobello20}, data reduction in physics \cite{Dellaquila21cpc} and medicine \cite{Buccheri21}, but has never been used to model nuclear physics data. Our data-driven method introduces a few key novelties in the study of nuclear fusion datasets: (i) the phenomenological model is derived exploiting a significant set of variables (we considered $25$ variables, including nuclear structure quantities such as charges, masses, proton and neutron separation energies, $\alpha$ $Q$-values, spin and parities of targets, projectiles and compound nuclei), thus allowing to perform a global search for computationally simple models to describe the data through a high-dimensional feature space; (ii) the so called \emph{feature selection}, i.e. the capability to suitably reduce the number of variables exploited by the model, thus helping to simplify the model; (iii) the fulfilment of physical constraints (i.e. vanishing fusion cross section at very large and very low center-of-mass energies).

The manuscript is organized as follows. In Sect.~\ref{sec:braiproject} we describe the dataset used to derive the models and we introduce some concepts of the neural-genetic programming adopted for this study. In Sect.~\ref{sec:analisi}, we discuss the models obtained from the data analysis by using different trade-offs between complexity and accuracy, and we compare them with the outcomes of other state-of-the-art phenomenological an theoretical models. Finally, in Sect.~\ref{sec:conclusioni}, we draw some conclusions and summarize the findings of our work.

\section{Dataset and methods}
\label{sec:braiproject}
\subsection{Dataset collection, splitting and pre-processing}
\label{subsec:preparation}
%%% Parte aggiunta da Marco e modificata da Daniele 23/01/2022 %%%
The approach adopted in this work belongs to a branch of artificial intelligence called \textit{machine learning} (ML). ML deals with the implementation of algorithms that are capable to automatically improve their \emph{performance} in the execution of a given \emph{task}, exploiting a suitable set of \emph{examples} (namely \emph{patterns}). In the present case, the task is that of predicting the complete fusion cross section between two nuclei, given the nuclear species and their collision energy, and the performance, as it will be discussed in the following paragraphs, is intimately linked to the accuracy of the prediction, i.e. to the discrepancy between predicted and experimental values of the fusion cross section. A similar paradigm is called \emph{supervised learning}, because the patterns used to derive the model comprise the expected output (i.e. the target fusion cross section value, for each individual pattern). On the contrary, in \emph{unsupervised learning} applications, patterns are unlabeled.

Results obtained using ML algorithms depend heavily on data. We can reasonably say that almost each data set is imperfect. For this reason, data set collection and its organization are the most crucial aspects in order to ensure the convergence towards a plausible model. For example, in data classification, one often deals with unbalanced data (\emph{class imbalance}). This is the typical case in level-zero triggers
for nuclear physics, where the events of interest are very rare. In other cases, one could have data that must be re-arranged to take
into account real distributions. Typical examples are the sex, age,
ethnicity adjustment of medical datasets (see e.g. \cite{Buccheri21}).

A quite conventional approach used in data set adjustment is \emph{pattern weighting} (see for example Ref.~\cite{Russo00}). This technique consists in adding a weight to each pattern belonging to the data set. This permits to modulate the error distribution among the patterns in a reasonable way, accounting for the experimental error associated to each pattern and/or increasing the importance of underrepresented classes.

As will be clearer in the following, our data set is very critical for several reasons: (i) cross section values range for several order of magnitude, and consequently, the corresponding uncertainties; (ii) there are collisions systems strongly underrepresented; (iii) relevant data for some regions in the search space are completely missing or poorly represented. To account for all these aspects, with the aim to develop simple
but plausible models, we performed a very accurate pre-processing of our dataset which consisted in $7$ individual steps.

The experimental data of fusion cross sections were extracted from the Nuclear Reaction Video Project database \cite{NRV2017,NRVonline}. In particular, we focused our attention to the cross section of fusion-evaporation events and to relatively light systems. In this way, we can neglect, in first approximation, the possible presence of fusion-fission phenomena, for which few data are available in the literature. In the following discussion, we will indicate with the subscripts $p$, $t$, $cn$ properties related to the projectile, target and compound nucleus  respectively, of a given collision.

\subsubsection{Data set splitting: light to medium mass systems and heavy ones.}
The database was built by considering systems with $Z_p\ge6$ and $Z_t\ge6$, and with $A_{cn}\le130$. In this way, we can neglect too heavy systems, for which fusion-fission can be the dominant reaction mode, and also too light systems, where the presence of break-up and transfer reactions complicate the analysis. A further selection concerns with the energy of data points: to avoid regions where the data can be influenced by the onset of \textit{incomplete} fusion \cite{Morgenstern84}, we considered reactions with bombarding energies smaller than $5$ MeV/nucleon.

The dataset was then divided into two groups: in the first one (preliminary \emph{learning set}), we required $Z_p\cdot Z_t < 250$; in the second one (first \emph{testing set}) we put the data with $Z_p \cdot Z_t \ge 250$. The models described in this paper are derived by using a part of the preliminary learning set, while the first testing set, for which one could expect the cross section being influenced by fusion-fission phenomena, was used to check the extrapolation of the resulting models towards heavier nuclear systems.

\subsubsection{Management of uncertainty: the first step of pattern weighting.}
At a first instance, experimental errors were used to build the statistical weight of each experimental point in the dataset. The weight of a given point was chosen as $1/\sigma_i^2$ \cite{Barlow89}, where $\sigma_i$ is the statistical uncertainty associated with the $i$-th point.

\subsubsection{Balancing collision systems: the second step of pattern weighting.}
Subsequently, we analyzed the total weights associated with each collision system. Not surprisingly, system weights resulted to be strongly unbalanced, i.e. some collision systems dominated the overall weight and most of other collision systems were critically underrepresented. This is due to the fact that complete fusion cross section data have large variations, as a function of energy, ranging from region 0 to region II, and the data of region 0 are characterized by even extremely small cross sections and correspondingly large weights. This causes collision systems for which experimental investigations extended down to near-barrier energies to be characterized by way larger weights compared to systems for which most data covered exclusively regions I-III. Using these unbalanced weights could lead to models particularly accurate in region 0, which might easily reach small total prediction errors, but with poor generalization capabilities in regions I-III.

In this framework, as in several ML applications, balancing pattern weights is therefore a necessary step for the learning phase, to effectively help generalizing the behavior of the cross section for different systems. Ultimately, the overall weight of each collision system was normalized to $1$ by scaling all weights within the system by a common factor. In this way, each collision system had the same importance in the learning phase. To avoid overweighting of systems with limited availability of data, we excluded from the learning set collision systems that had less than $5$ data points and, simultaneously, a weight lower than $5\%$ of the sum of all weights in the dataset (before re-scaling them). However, these collisions systems are not excluded from the overall derivation of the model: they will represent a powerful benchmark to inspect the prediction capabilities during the testing phase of the models, as it will be discussed in the following paragraph.

\subsubsection{Second phase data set splitting.}
An useful testing set for ML approaches should contain data, not included in the dataset used for model learning, which suitably covers the relevant feature space. For this reason, alongside with heavy collision systems, excluded from the preliminary learning test and included in the first testing set to check the extrapolation capabilities of the models towards heavier systems, some of the points belonging to the preliminary learning set were selected and included in the second testing set. In addition to these points, the second testing set also included experimental points belonging to collision systems excluded from the learning set in the second step of pattern weighting, as described in the previous paragraph. This is a crucial aspect of our analysis as it allows to effectively probe the generalization capabilities of our models by analyzing their extrapolations to other light to medium mass nuclei besides those used for the learning phase. All remaining points, which were not included in the second testing set, constitute the final learning set.

\subsubsection{Database enrichment in poorly represented regions trough artificial data.}
An inspection of the experimental data reported in the literature clearly shows a lack of data in the region III (see Fig.~\ref{fig:all_systems_0}). Since it is important to have reference points in this energy regime to try a description of the whole trend of fusion data, we enriched the database by considering the comprehensive analysis on region III data made by Eudes et al.~\cite{Eudes14}. To this end, we used the predictions of the homo-graphic function of \cite{Eudes14} to generate, for each system, $N_{high}$ points randomly distributed in such high energy region. $N_{high}$ was treated as an adjustable parameter. The boundaries of region III were reasonably chosen between $1.5$ and $6$ MeV/nucleon for light systems ($A_{cn}<80$) and between $3.5$ and $6$ MeV/nucleon for heavier systems ($A_{cn}\ge80$). In the analysis, such points had a weight corresponding about $15\%$ of the total for a given system, and optimal solutions were typically obtained with $N_{high}=10$.

\subsubsection{Fulfillment of physical boundaries.}
An important point of the approach implemented in this study concerns with the presence of physical boundaries, i.e. predicted cross section values should drop to zero at considerably low and considerably high energies, respectively because of the huge effect of Coulomb barrier and of the onset of multi fragmentation mechanisms. To fulfill such requirements, we introduced 2$N_r$ \textit{regularization points} in the database, for each system. These points, having zero value of the cross section, are randomly placed both at negative and at large values of $1/E_{cm}$ and account for $5\%$ weight with respect to the total weight of a given system. $N_r$ was treated as an adjustable parameter; in our analysis, we found that $N_r=10$ was an optimal choice to achieve satisfactory boundaries for the derived models.

\subsubsection{Filtering outliers.} 
If we globally inspect the data after applying the constraints described above, by plotting them (as usual) in the $1/E_{cm}$ - $\sigma_f$ plane, we can often observe the presence of some \textit{outlier} points, that visually disagree with nearby ones. Since such points could bias and worsen the outcomes of the neural-genetic algorithm, we implemented a suitable \textit{filter} to discard them from the analysis. Points were filtered by accounting for the Euclidean distance $d$ between pairs of points in the $1/E_{cm}$ - $\sigma_f$ plane. For a given point, we search and \textit{count} its \textit{nearby points}, i.e. those points having a distance lower than a given threshold value $d_{thr}$. If the number of nearby points was found to be smaller than a predefined fraction $f_{near}$ of the total number of points considered for a given system (obviously excluding the simulated points in region III and the regularization points), then that point was defined as \textit{isolated} and discarded from the dataset. This filtering procedure was applied iteratively \textit{two times} to the data: after a first filtering, in fact, some non-isolated points could become isolated. The parameters were chosen to be $d_{thr}=0.06$ and $f_{near}=4\%$, after dedicated tests. We verified that the use of this protocol represents the best compromise between an excessive reduction of data and an excessive scattering of them.
%%% Fine parte aggiunta da Marco %%%%%%%%%%%%%%%%%%%%%%%%%
%%%%%%%%%%%%%%%%%%%%%%%%%%%%%%%%%%%%%%%%%%%%%%%%%%%%%%%%%%

\subsection{Description of the neural-genetic algorithm: the Brain Project}
To derive the models discussed in this work, we used the Brain Project (BP), a state-of-the-art tool for the formal modeling of data based on a hybridization of genetic programming and artificial neural networks. The implementation details of BP are deeply discussed, for example, in Refs.~\cite{Russo16,Russo20}, and go beyond the purpose of the present paper. In this section, we describe, in a schematic way, the main aspects of the underlying mechanism.

BP implements a programming technique inspired by the natural selection in biological systems (see, e.g. \cite{Cooper97}) called genetic programming (GP). GP is part of the field of Evolutionary Computing \cite{Koza92}. Within this field, a line of research particularly interesting for studies involving datasets of physics data is that named \textit{Symbolic Regression}, which is the process during which some data are fitted by means of a suitable analytical formula. Unlike traditional fit of data, in Symbolic Regression algorithms the analyitical expression of the formula used for the fitting is derived by the algorithm itself. With this respect, many novel approaches have shown that the evolutionary core of GP has the capability to deal with extremely complex problems. In addition, some GP implementations can also perform the so-called \emph{feature selection}, i.e. they are capable to suitably exclude some of the variables used as input according to some predefined \textit{trade-off} between accuracy and complexity of the model. The latter is a crucial aspect in the modeling of nuclear data, as it allows to investigate the existence of correlations between variables without resulting in too complex models. Furthermore, this aspect is particularly important when the cardinality of data to fit is significant.

In the approach implemented by BP, applied to the present case, a number of models, each one representing a particular function used to reproduce phenomenologically the cross section values of the entire dataset, are randomly generated. They form the \emph{individuals} of a so-called \textit{population}. The size of the population, i.e. the number of its individuals, is constrained to be \textit{constant}, as typically occurs in biological systems at the saturation of a logistic growth \cite{CurtisHelena1989}. Each individual is then associated with a so-called \emph{fitness}, which quantifies the goodness of an individual with respect to the problem of fitting the input data (in this case, the nuclear fusion cross section data). The following operations are then iterated, until a predefined convergence criterion is reached:
\begin{enumerate}
	\item Inside a given population, two individuals are selected (typically \textit{randomly}, or by using stochastic or deterministic criteria based on their fitness) and used to generate an \textit{offspring}. This latter inherits part of the \emph{chromosomes} of the \emph{parents}, i.e. \textit{functions}, \textit{operations} and \textit{constants}. The inheritance of such a genetic material follows a stochastic coupling similar to the \textit{crossing-over} process during the meiosis in biology \cite{Cooper97}, helping to increase the \textit{variety} in the population.
	\item with a certain probability, a \textit{genetic mutation} occurs (i.e. low probability changes in the type of function, operation or constant that might appear in a newly generated individual before it is included in the population).
	\item Since the size of the population is kept constant, the newly generated offspring replaces a previously existing individual in the population. Even if, stochastically, there is a non-vanishing probability for each individual in the population to be replaced by a new individual, the individual to be replaced is typically chosen following a \textit{probability} distribution which accounts for its \textit{fitness}, i.e. low-fitness individuals might have greater probability to be replaced. This process is similar to that happening in \textit{Darwinian evolutionism} with the so called \textit{natural selection} \cite{CurtisHelena1989}: only individuals whose attitudes and skills fit well the environmental context will have \textit{a chance} to survive and to transmit their genetic patrimony to sons.
	\item The fitness $f_{fit}$ of the new offspring is calculated.
\end{enumerate}
In the end, the best individual, i.e. the tree structure (see Section 2.2.1) with the maximum value of $f_{fit}$, is designated as the output result.

With this evolutionary scheme, one could imagine to easily find a convergence, i.e. to find, after several iterations, a population in which there is an individual characterized by leading to a small \textit{prediction error} $e$, calculated in the dataset used to derive the model. In the framework of BP, the prediction error is defined as:
\begin{equation}
\label{eq:error}
e=\frac{100}{\sigma_y}\sqrt{\Sigma_{i=1}^{N_{p}}\frac{w_{i}(y_i-y_i^{th})^2}{\Sigma_{j=1}^{N_p}w_j}}
\end{equation}
being $y_i^{th}$ the target output value expected for the $i$-th point (i.e, the experimental complete fusion cross section of the $i$-th point in the dataset), $w_i$ its weight, $y_i$ the corresponding value predicted by the model, $N_p$ the number of patterns in the learning set, and $\sigma_y$ is the weighted standard deviation of all the $y_i^{th}$ values in the learning set.

Unfortunately, some problems can occur in the mechanism of evolutionary phase. For example, one could be in the presence of a so-called \textit{superindividual}, i.e. an individual characterized by a large fitness value compared to the competitors in the population, even if being located in a relative minimum of the prediction error. This unfavorable situation is often referred as \textit{premature convergence}, and can result in the population becoming \textit{stagnant}. To mitigate this problem, several strategies are usually adopted. Another crucial aspect in GP is the \textit{migration of individuals} from different populations. In analogy with biological systems \cite{CurtisHelena1989}, this also would help to avoid stagnation due to the supply of new genetic material. It is worth noting that the implementation of such a migration requires the simultaneous presence of more than one population. This makes GP particularly suitable for parallel and distributed architectures, especially in multithreading machines.

Finally, the neural part of BP is a relatively time wasting task. This is invoked only when one individual is particularly promising. To this end, a mathematical expression is suitably transformed into a multi-layer feed-forward neural network by replacing operators in the original expression with specialized neurons, while all constants are considered as weights of the artificial neural networks. The mathematical expression of the error gradient in the weight space is then calculated and an enhanced steepest descending technique is used to update the weights. Separate learning rates for each weight are dynamically adjusted during the training.

\subsubsection{Characterization of an individual and definition of the fitness function}
\begin{figure}
	\centering
	\includegraphics[scale=0.5]{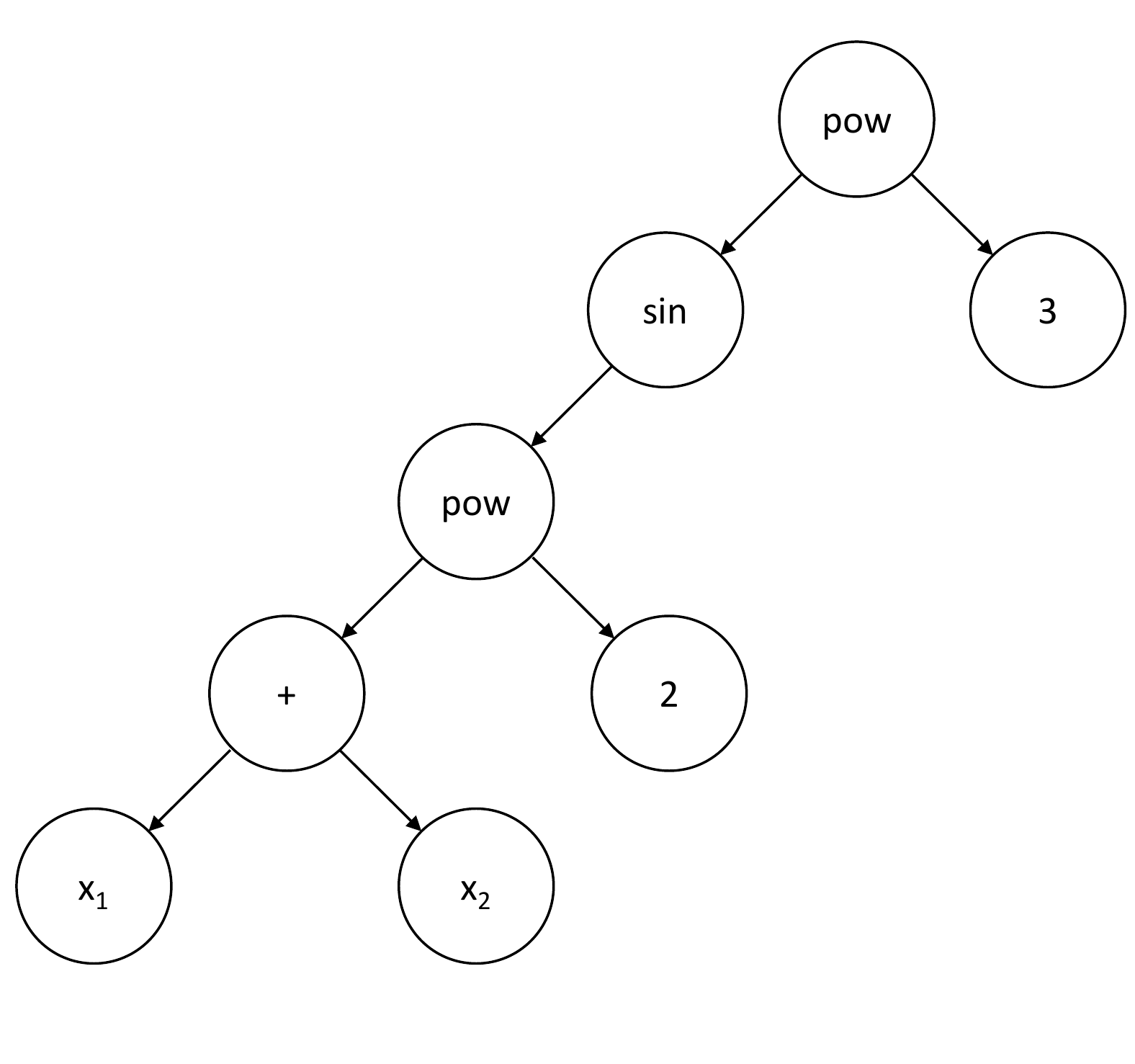}
	\caption{\label{fig:tree_structure} The tree-like representation of the expression $\sin^3\left[(x_1+x_2)^2\right]$.}
\end{figure}
The BP approach is part of a line of GP research that foresees the evolution of tree structures representing formal mathematical expressions. In such tree-like structures, branches are interconnected by \textit{nodes}. A \textit{node} can represent an \textit{operation} (e.g. $+,-,\times,\div,\wedge, etc$), a \textit{function} (e.g. $\sin$, $\cos$, $\exp$, $\log$, $\erf$, and similar common functions), a variable, or a \textit{numerical value} (constant). As an example, Fig.~\ref{fig:tree_structure} shows the tree-like representation of the expression $\sin^3\left[(x_1+x_2)^2\right]$. $x_1$ and $x_2$ are the variables used for the calculation of the function (usually called \textit{features}). In this representation, it is evident from the figure that each individual assumes a \textit{tree-like} shape, where the nodes are indicated as circles. We can easily recognize that this expression comprises $8$ nodes. In BP, the number of nodes of an expression is used to quantify the computational complexity of the formula, i.e. an expression foreseeing a lower number of nodes is considered less mathematically complex.

The fitness function $f_{\rm fit}$, which is greater than or equal to zero, accounts for the prediction error ($e$, defined by eq.~\ref{eq:error}), the number of nodes and the number of features exploited by the model. BP maximizes $f_{\rm fit}$ in the learning phase. In the current implementation of BP, several operational modes are possible. In the simplest, BP is used to minimize the prediction error without additional requirements. A more complex mode involves the definition, by the user, of a target error $e_{tgt}$. In this mode, BP searches for the simplest model with a prediction error lower than or equal to $e_{tgt}$. An additional mode, which was largely used in the present study, is that in which a target number of nodes $n_{tgt}$ is defined by the user and BP searches for the most accurate model (which minimizes the error $e$) but with a number of nodes lower than or equal to $n_{tgt}$. A similar mode exists also with a target number of features $f_{tgt}$, and can be used to try to reach models with $f_{tgt}$ features. A complete description of the fitness function used by BP can be found, for example, in Refs.~\cite{Russo16, Campobello20} and goes beyond the scope of the present work.

\section{Fusion data analysis}
\label{sec:analisi}
\begin{figure}
	\centering
	\includegraphics[width=0.6\textwidth]{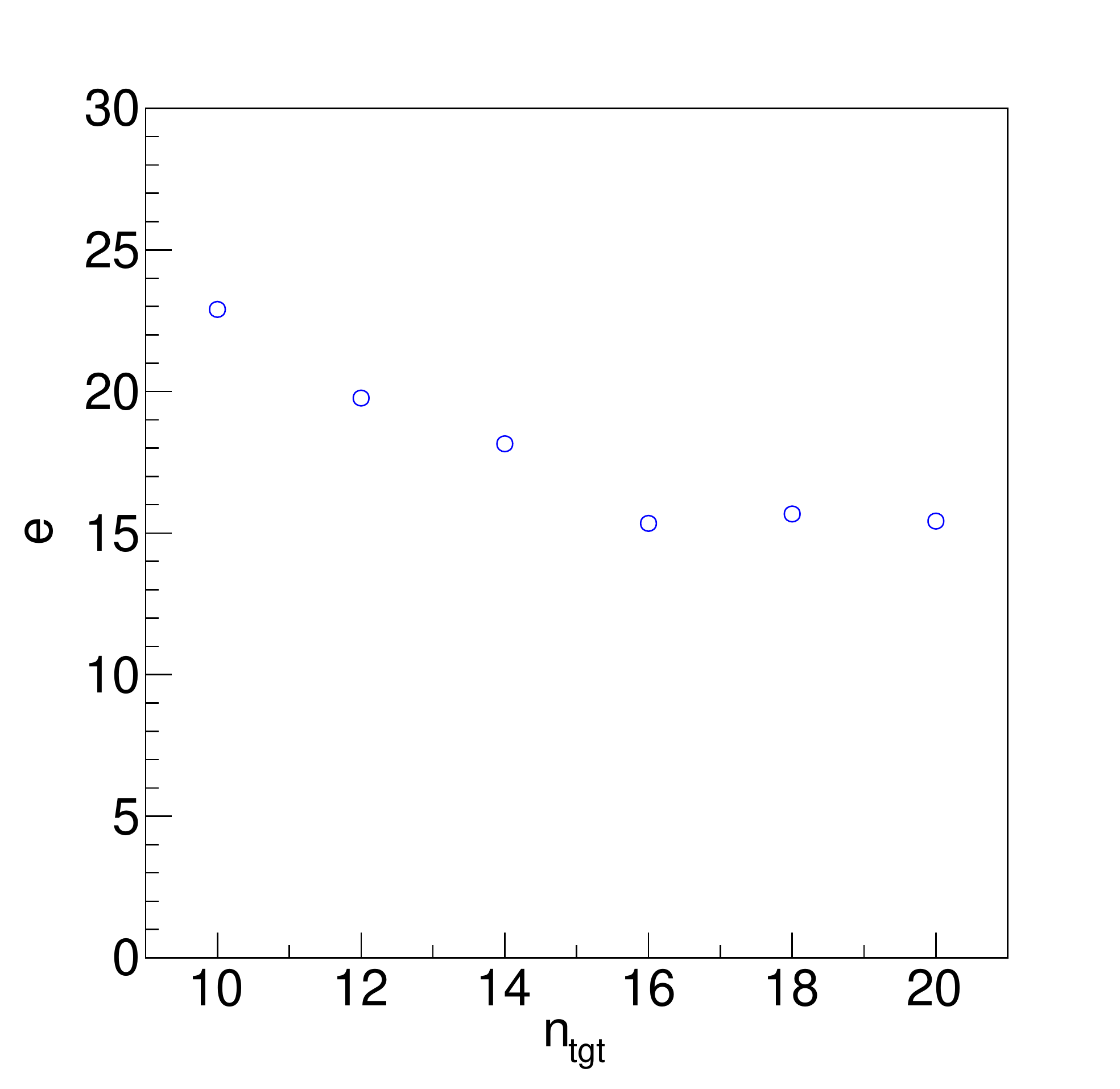}
	\caption{\label{fig:fitness} Trend of the best error value $e$, as a function of the number of target nodes $n_{tgt}$. More details on the definition of $e$ are described in the text.}
\end{figure}

We fitted the complete fusion data belonging to $124$ light to medium mass systems taken from the database of Refs.~\cite{NRV2017, NRVonline}, by using the instructions described in Sect.~\ref{sec:braiproject}. The dataset used to derive the models ultimately contained about $4500$ experimental points. Data associated with $63$ medium-mass systems (with $Z_p\cdot Z_t>250$) were also used to check the extrapolations of the obtained phenomenological formulas to heavier systems, but such points were not used in the derivation of the models, as explained before. Besides the variation of the experimental complete fusion cross section $\sigma_f$ with the inverse of the center of mass energy $1/E_{cm}$, we considered also the possible effect of other $24$ variables characterizing the structure of projectiles, targets and compound nuclei. We used in turn $25$ numerical \emph{features}, which are summarized in Table~\ref{tab:features}, alongside with their meaning.
\begin{table}[t]  
	\centering
	\begin{tabular}{cc}
		\hline
		Symbol                              & Description               \\
		\hline
		$\frac{1}{E_{cm}}$                  & inverse of the collision center-of-mass energy (MeV)                  \\
		$Z_1$                  & charge of the projectile                   \\
		$Z_2$                  & charge of the target                   \\
		$A_1$                  & mass of the projectile                   \\
		$A_2$                  & mass of the target                  \\
		$J_1$                  & spin of the projectile                   \\
		$J_2$                  & spin of the target                  \\
		$\pi_1$                  & parity of the projectile ($1$ for positive parity, $-1$ for negative parity)                   \\
		$\pi_2$                  & parity of the target ($1$ for positive parity, $-1$ for negative parity)  \\
		$\mu_1$                  & magnetic dipole momentum of the projectile ($\mu_N$)              \\
		$\mu_2$                  & magnetic dipole momentum of the target ($\mu_N$)                 \\
		$\langle r^2\rangle_1$                & rms charge radius of the projectile (fm)              \\
		$\langle r^2\rangle_2$                & rms charge radius of the target (fm)                 \\
		$Q$-value                & fusion $Q$-value (MeV)                 \\
		$S_\alpha$                  & $\alpha$ separation energy of the compound nucleus (MeV)                   \\
		${S_\alpha}_1$                  & $\alpha$ separation energy of the projectile                  \\
		${S_\alpha}_2$                  & $\alpha$ separation energy of the target                  \\
		${S_n}_1$                  & one-neutron separation energy of the projectile                \\
		${S_n}_2$                  & one-neutron separation energy of the target                \\
		${S_p}_1$                  & one-proton separation energy of the projectile                \\
		${S_p}_2$                  & one-neutron separation energy of the target                \\
		${S_{2n}}_1$                  & two-neutron separation energy of the projectile                \\
		${S_{2n}}_2$                 & two-neutron separation energy of the target                \\
		${S_{2p}}_2$                  & two-proton separation energy of the projectile                \\
		${S_{2p}}_2$                  & two-neutron separation energy of the target                \\
		\hline
	\end{tabular}
	\caption{List of the features used to train the models. They are selected to suitably represent each collision system and the relative energy between projectile and target of each data point. After model derivation, the simplest model foresees exclusively $3$ features while the most complex model exploits $4$ features.}
	\label{tab:features}	
\end{table}

We performed several trials with a different number of target nodes $n_{tgt}$ at increasing values starting from $n_{tgt}=4$. The optimal values of the error function (as well as the fitness function) were found in the range $n_{tgt}\simeq 10-20$, showing clear saturation effects starting from $n_{tgt}=16$, as shown in Fig.~\ref{fig:fitness}. This fact suggests that, even increasing the complexity of the model, it is not possible to gain a statistically significant improvement of the prediction error. Furthermore, for each $n_{ngt}$ value, we verified that the prediction error on the second testing set (which covers the relevant feature space of the learning set) obtained with the resulting model, was always of the same order of that obtained for the learning set, thus ensuring the absence of \emph{overfitting} phenomena even for the most complex models. Besides overfitting phenomena, which are extremely unlikely to occur due to the limited number of nodes and the complexity of the dataset, the second testing set was also used to test the generalization capabilities of the models. This is a crucial aspect in physics. A suitable model to predict the fusion cross section must have generalization capabilities, which allow its extrapolation even to new collision systems. The second testing set contained several light to medium mass collision systems that were completely unrepresented in the learning set. For each of those systems, and regardless the number of target nodes, we obtained similar prediction errors than in the systems included in the learning set, thus testifying the good extrapolation capabilities of the proposed models.

Guided by these initial findings, we propose two different phenomenological formulas to describe the complete fusion cross section of light to medium mass nuclei, with two different trade-offs between complexity and accuracy. The first formula is extremely simple and handful, and therefore particularly suitable to be used for fast evaluation of the expected complete fusion cross section of a given system at given energy. The simplest model exploits only $3$ features, namely $1/E_{cm}$, $Z_1$ and $Z_2$ and $10$ nodes, and has the following expression:
\begin{equation}
	\sigma_{fus}^{(n_{tgt}=10)}(E_{cm})={1102}\cdot {\exp\left[-\left({1.39}{{-0.469}\cdot {Z_2}\cdot {Z_1}\cdot {\frac{1}{E_{cm}}}}\right)^2\right]}
\end{equation}
in unit of mb. The second formula, analytically more complicated but with a better accuracy, is obtained with $n_{tgt}= 20$ and has the following expression:
\begin{equation}
	\begin{split}
	\sigma_{fus}^{(n_{tgt}=20)}(E_{cm})&={144.5}\cdot{\erf\left\{\exp\left[-\left({0.129}\cdot {\frac{1}{E_{cm}}}	     \cdot {Z_1}\cdot {\left({A_2}+{\ln\left({S_{2n}}_2\right)}\right)}\right)^2\right]\right\}}\cdot\\
	&\cdot{e^{\erf\left({102.3}\cdot {\frac{1}{E_{cm}}}\right)}}\cdot {Z_1}\cdot {\frac{1}{E_{cm}}}\cdot {A_2}
	\end{split}
\end{equation}
where the features $1/E_{cm}$, $Z_1$, $A_2$ and ${S_{2n}}_2$ are used.

A first interesting comment can be made observing the features selected by BP. Using the present complete fusion data available in the literature and the dataset discussed in Sect.~\ref{sec:braiproject}, the neural-evolutionary algorithm was able to find strong correlations of the fusion cross section mainly with the product $Z_1Z_2$ (or with similar quantities, as $Z_1A_2$) and the inverse of the center of mass energy $1/E_{cm}$. This fact is not surprising, considering that the fusion process is intimately linked to Coulomb barrier. Side effects of the Coulomb related term $Z_1Z_2$ on the fusion cross section in heavy systems were pointed out, for example, by Vaz et al.~\cite{Vaz87}, while it is worth noting that several other phenomenological approaches show a clear dependence on $Z_1Z_2$ (e.g., \cite{Porto84,Giordano90,Lozano80,Kailas81}). Besides the very weak, logarithmic, dependence on the target two-neutron separation energy present in the higher complexity model, no significant correlations on other features related to nuclear structure properties, or even mass asymmetries, can be evidenced with this approach. This is a particularly interesting fact as it suggests that the informative content of present nuclear fusion datasets might be essentially contained in the set of variables $E_{cm}$ and $Z_1Z_2$ (or $Z_1A_2$). For this reason, the inclusion of additional variables might not add further information to significantly improve the description of current experimental data.

The functional forms obtained with our approach are quite simple. The trend is fully described by a Gaussian function for the model with $n_{tgt}= 10$, and by a combination of error functions, exponential and linear functions in the more complicated case of $n_{tgt}= 20$. Despite its simplicity, the $n_{tgt}= 10$ formula allows some quick estimate of fusion cross sections in the regions I and II, but it becomes unreliable at very large energies in region III, where it predicts small but non-vanishing fusion cross sections. The $n_{tgt}= 20$ formula reproduces better the trend of data in region III and matches the $\sigma_f=0$ constraint at high energies, even if its analytic behavior, once transformed in the $\left(\ell_{cr},E_x\right)$ plane (see, e.g. Refs.~\cite{GomezDelCampo73,Beck96,Cavallaro98}) does not show the clear occurrence of a limiting critical angular momentum. It is also worth nothing that, despite our approach is completely data-driven, resulting models are symmetric (for $n_{tgt}=10$) or nearly-symmetric (for the most complex model) to the kinematics of the entrance channel (i.e. to projectile-target exchange). This is an interesting finding as the symmetrical physics requirement is here uniquely suggested by learning data, without assuming additional constraints\footnote{Even allowing for larger complexity models, BP suggests kinematically symmetric $\sigma_f$ functions. Only the $n_{tgt}= 20$ formula shows small asymmetries, comprising the product $Z_1A_2$ instead of $Z_1Z_2$ or $A_1A_2$, excluding the logarithmic dependence on $S_{2n2}$.}.

\begin{figure}
	\centering
	\includegraphics[scale=0.8, angle=0]{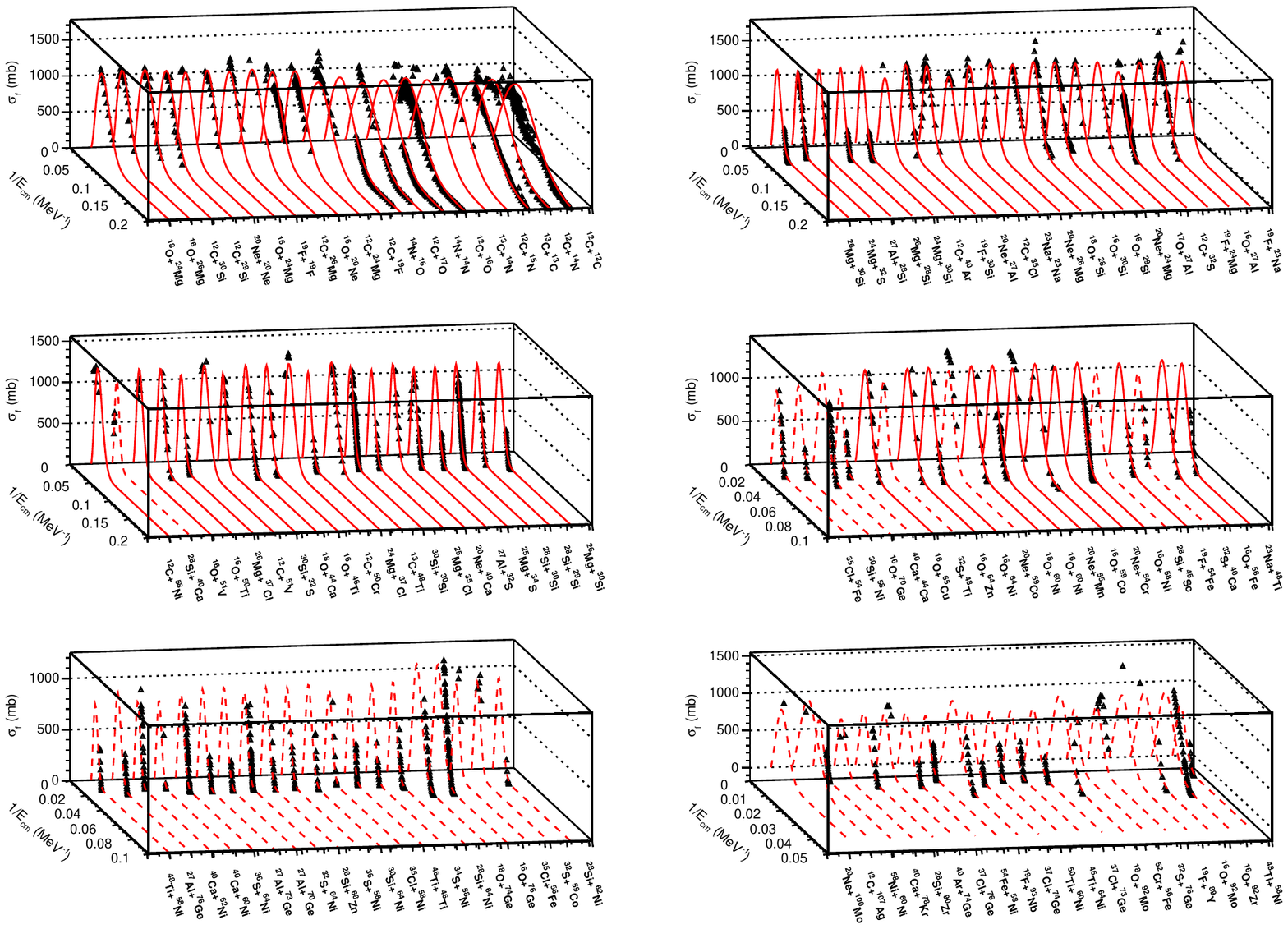}
	\caption{\label{fig:all_systems_0} Fusion cross section data derived from the NRV database \cite{NRV2017,NRVonline} for several light systems with $A_{tot}>24$ (black dots with error bars). For clarity reasons, we report only a random subset of the collisions systems explored here. Red solid lines are the results of the phenomenological formula with $n_{tgt}=20$. Red dashed lines show the results of the $n_{tgt}=20$ for systems included in the first testing dataset. They are useful to test the extrapolation capabilities of the model towards heavier systems.}
\end{figure}

\begin{figure}
	\centering
	\includegraphics[width=\textwidth]{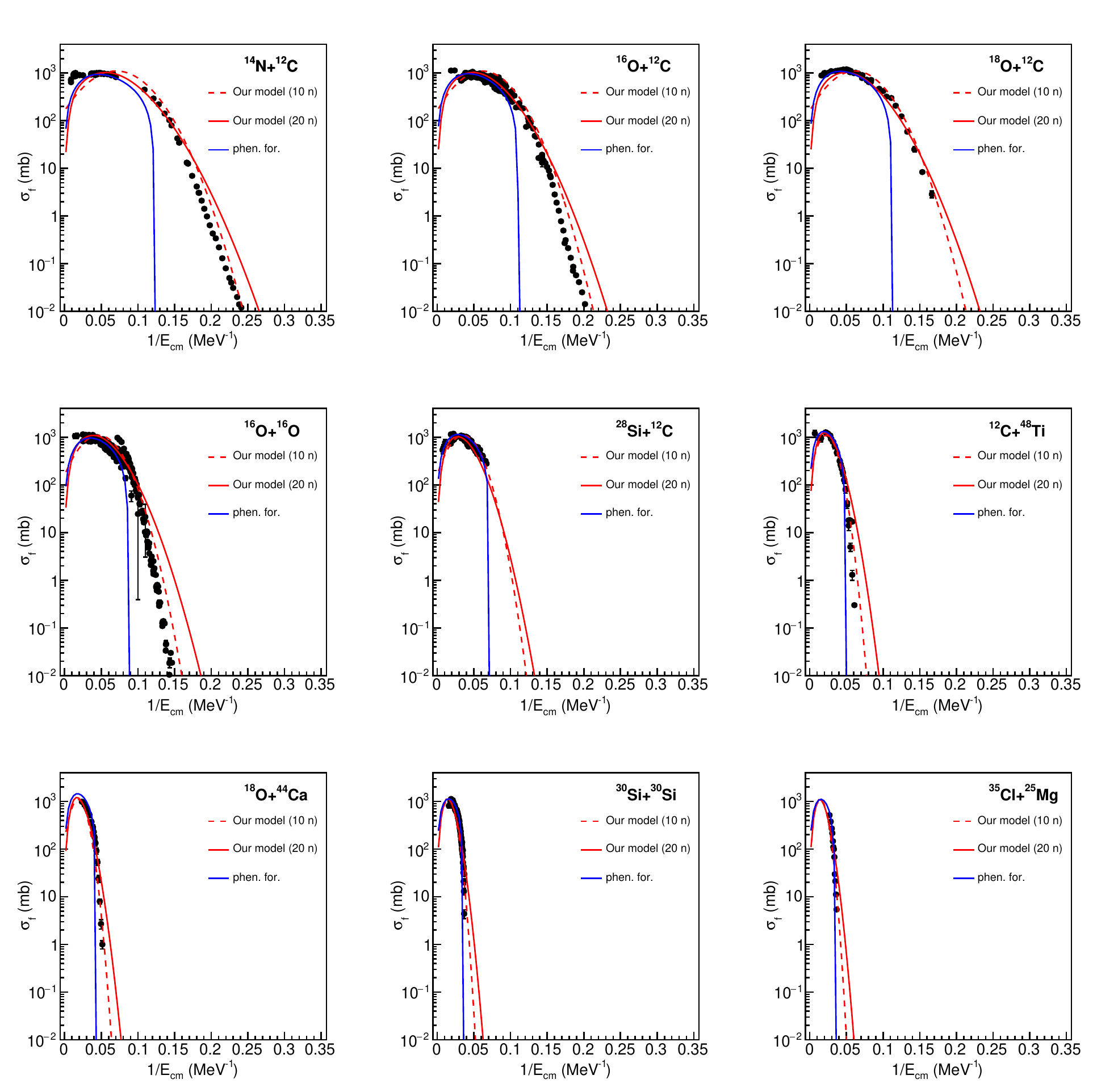}
	\caption{\label{fig:selected_systems} A close inspection of the results of our model compared to the experimental data for some selected light and medium mass systems. Experimental data are taken from the NRV database \cite{NRV2017,NRVonline}, and are shown as black dots with error bars. Red solid lines: our phenomenological formula results obtained with $n_{tgt}=20$. Red dashed lines: our phenomenological formula results obtained with $n_{tgt}=10$. Blue solid lines: results of the phenomenological formula obtained in Ref.~\cite{Porto84}.}
\end{figure}

\begin{figure}
	\centering
	\includegraphics[width=0.7\textwidth]{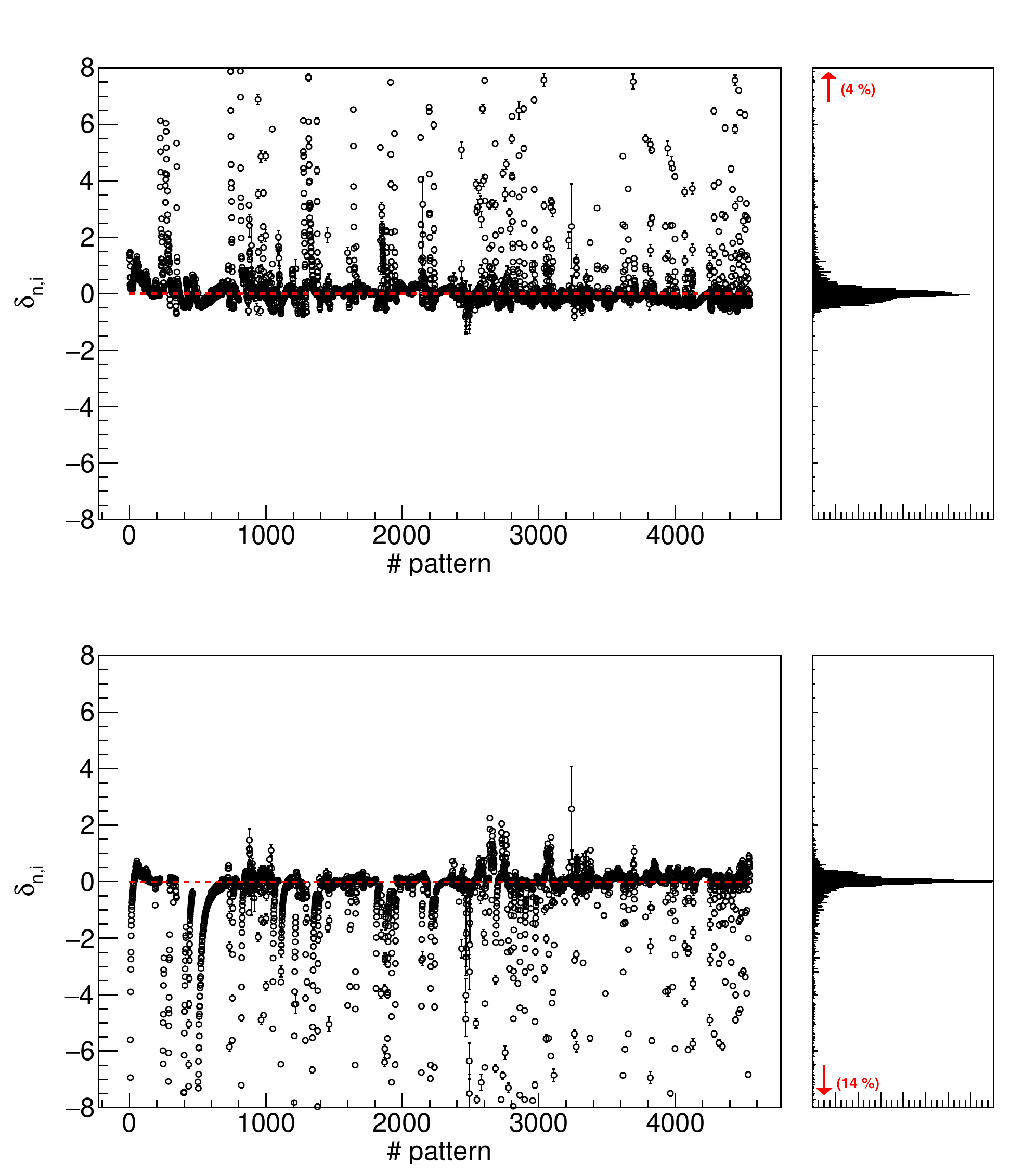}
	\caption{\label{fig:summary_light_systems}. (Upper panel) Differences between the predicted and experimental values of the fusion cross sections obtained for systems with $Z_1Z_2<250$, normalized to the experimental ones. Predictions are obtained with the phenomenological formula derived in this work with $n_{tgt}=20$. In the lateral insert, we show the projection of the differences on the vertical axis. The arrows and the number in parenthesis indicate the percentage of points outside the graph with respect to the total number of points. (Lower panel) As in the upper panel, but here the predictions are obtained with the phenomenological formula discussed in Ref.~\cite{Porto84}.}
\end{figure}
In Figure \ref{fig:all_systems_0}, we show the trend of the phenomenological formula with $n_{tgt}= 20$ with respect to the experimental data for all collision systems in the database. Solid lines refer to the lightest systems ($Z_1Z_2<250$), which were used to derive the models. Dashed lines show the extrapolation of our model to systems with higher masses, excluded from the training phase and used to test the validity of the model when extrapolated towards heavier systems (first testing set). We can see a reasonably good description of the whole database of light systems, especially in the regions I and II of the excitation functions\footnote{It is interesting to note that the artificial intelligence approach suggested the presence of some typos (concerning the labels related to the use of the CM or LAB frames for the fusion data) in the NRV database. In detail, $^{16}\textrm{O} + {^{46}\textrm{Ti}}\rightarrow {^{62}\textrm{Zn}}$, $^{16}\textrm{O}+ {^{64}\textrm{Zn}}\rightarrow {^{80}\textrm{Sr}}$ and $^{16}\textrm{O}+ {^{66}\textrm{Zn}}\rightarrow {^{82}\textrm{Sr}}$ complete fusion cross section data extracted from Ref.~\cite{Gomes89} are incorrectly labeled with the energy in the laboratory reference frame; correct energy labels should be in the center-of-mass frame.}. Looking at the extrapolation of the results to heavier mass systems (dashed lines), a reasonable agreement is seen for data in region I; however, we remind that the proposed model was focused and tared on light to medium mass collision systems.

The main discrepancies between the model predictions and the data concern the region 0 of systems with large masses, i.e. the sub-barrier energy domain of heavy systems. In fact, it is generally difficult to reproduce, with a unique function valid at all energies, the sudden drop of the cross section due to the huge effect of barrier penetration in heavy ion fusion and to the impact of deformations and shell effects, taking also into account that the lower limit of $100$ $\mu$b used in the data fitting is quickly reached in heavy systems. Furthermore, we remind that the formula is derived by using, in the learning phase, exclusively light to medium mass system.

In Fig.~\ref{fig:selected_systems}, we compare our results, for some light systems, with the prediction of a phenomenological approach adopted to describe fusion data \cite{Porto84}, previously published in the literature. This model is derived starting from the modified sum-of-difference method (MSOD), and it can also reasonably reproduce some details in proximity of the maxima of fusion excitation functions. As evident, the neuro-genetic algorithm achieves a better description of the data at large $1/E_{cm}$ values for light systems, where the model of Ref.~\cite{Porto84} falls off too rapidly. Nevertheless, our approach and the phenomenological one predict similar results in the region of the maximum of the cross section. It is interesting to note that, for heavier systems, the predictions of \cite{Porto84} are closer to data in the region 0 of the excitation function. This effect is due to freedom of the phenomenological formula to go towards negative values of the cross sections at large $1/E_{cm}$ values, so being able to reproduce the sudden decrease of the heavy ion fusion cross section due to the Coulomb barrier penetration.

The accuracy of the two different approaches in describing the fusion cross section data can be quantitatively compared by inspecting the relative deviations $\delta_{n,i}$ between the data and the model predictions. We define the normalized deviations as:
\begin{equation}
\delta_{n,i}=\frac{\sigma_{f,i}^{pred}-\sigma_{f,i}^{exp}}{\sigma_{f,i}^{exp}}
\end{equation}
where $\sigma_{f,i}^{pred}$ and $\sigma_{f,i}^{exp}$ are the predicted and observed fusion cross sections of the $i$-th data point in the dataset, respectively. In Fig.~\ref{fig:summary_light_systems}, we show the calculated $\delta_{n,i}$ values for all data points of light collision systems ($Z_1Z_2<250$) obtained with the higher complexity model derived via BP ($n_{tgt}= 20$, upper panel) and with the phenomenological formula of \cite{Porto84} (lower panel). In both cases, the bulk of data is close to predictions, with $\delta_n$ mainly inside $\pm 0.75$ range. Outside that range, the present approach has a tendency to overestimate the data, while the phenomenological formula of \cite{Porto84} tends to underestimate the data. In our case, anyway, the percentage of outliers (with $|\delta_n|>8$) is smaller than in the Ref.~\cite{Porto84} case ($4\%$ \textit{versus} $14\%$).

\begin{figure}
	\centering
	\includegraphics[width=0.8\textwidth]{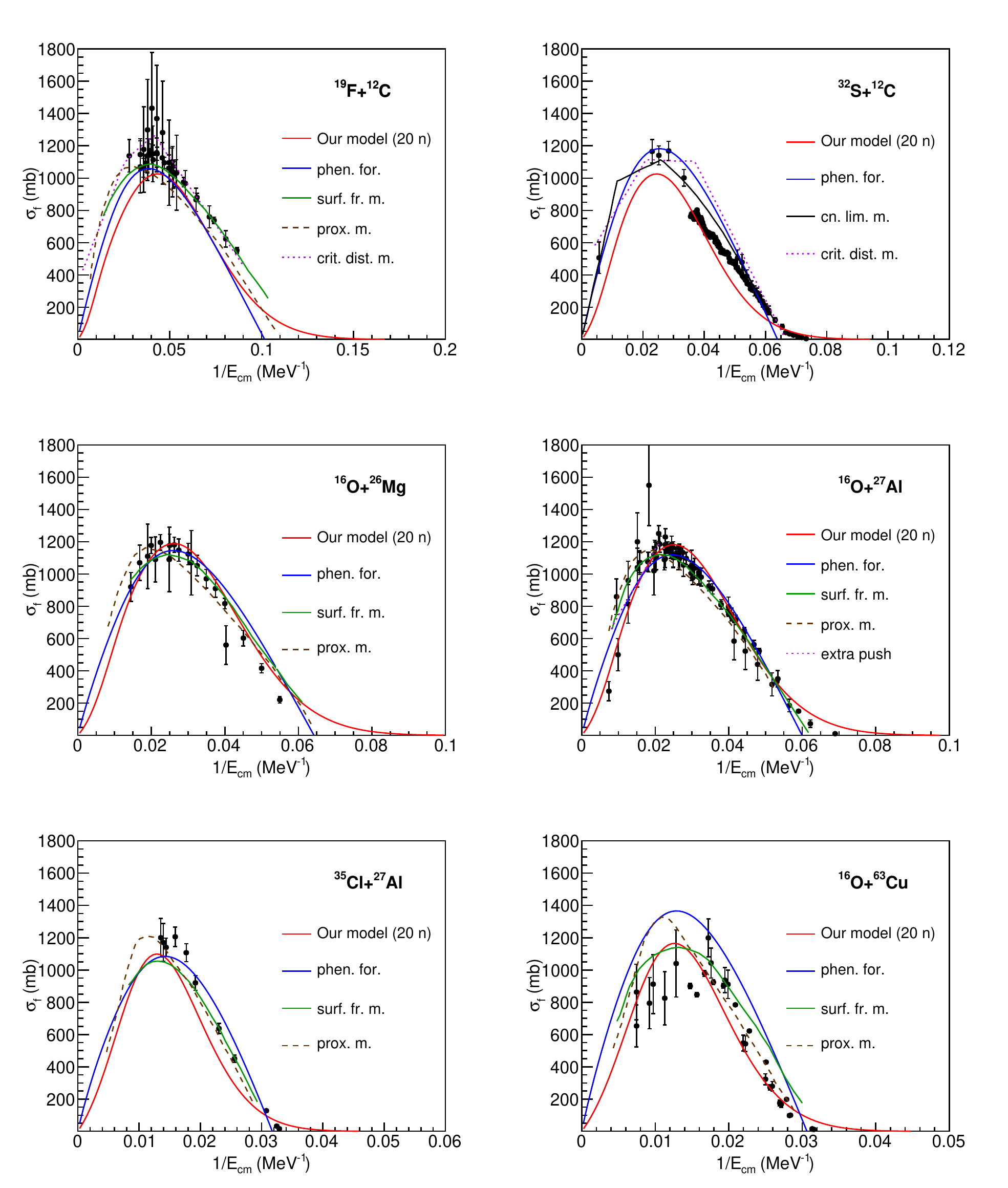}
	\caption{\label{fig:altrimod}. Fusion cross section for some selected systems in the $A_{tot}\simeq30-80$ domain. Data are shown in black dots with error bars; we included both direct and inverse kinematics data. Predictions of the newly proposed model are shown as red solid lines, while calculations with the phenomenological model of Ref.~\cite{Porto84} are shown as solid blue lines. Calculations with the surface friction model of Ref.~\cite{Frobrich84} are shown in green, while calculations performed with the proximity model \cite{Birkelund79} are shown as dashed brown lines. For the lighter systems, we show also the predictions of the critical distance model \cite{Lee80,Matsuse82} and of the limitation to the compound nucleus model by Bass~\cite{bass77}. ${^{19}\textrm{F}}+{^{12}\textrm{C}}$: data taken from \cite{Anjos90,Kovar79,Sperr76,Kohlmayer77,Puhlop75,Anjos02} in NRV and from \cite{Pop00}. ${^{32}\textrm{S}}+{^{12}\textrm{C}}$: data taken from \cite{Menacha90,Kolata85} in NRV and from \cite{Arena91,Pirr01}. ${^{16}\textrm{O}}+{^{26}\textrm{Mg}}$: data taken from \cite{Tabor78,Horn78} in NRV. ${^{16}\textrm{O}}+{^{27}\textrm{Al}}$: data taken from \cite{Rascher79,Kozub75,Kovar79,Eisen77,Lee81,Dauk75,Back77,Kowalski68,Gilfoyle92} in NRV. ${^{35}\textrm{Cl}}+{^{27}\textrm{Al}}$: data taken from \cite{Scobel76} in NRV. ${^{16}\textrm{O}}+{^{63}\textrm{Cu}}$: data taken from \cite{Langevin76,Natowitz70,Chamon92,Pereira89} in NRV.}
\end{figure}

Finally, the predictions of the newly proposed phenomenological formula and the phenomenological formula of Ref.~\cite{Porto84} can be compared to those of nuclear models attempting to describe the fusion between light to medium mass nuclei (Fig.~\ref{fig:altrimod}). To this end, we selected the systems ${^{19}\textrm{F}}+{^{12}\textrm{C}}$, ${^{32}\textrm{S}}+{^{12}\textrm{C}}$, ${^{16}\textrm{O}}+{^{26}\textrm{Mg}}$, ${^{16}\textrm{O}}+{^{27}\textrm{Al}}$, ${^{35}\textrm{Cl}}+{^{27}\textrm{Al}}$ and ${^{16}\textrm{O}}+{^{63}\textrm{Cu}}$, having total masses in the region $A\approx30-80$ and for which previous experiments reported data points in energy regimes ranging from region II up to, for some of the cases, region III. It is evident that our model tends to overestimate data in the region 0, while it reasonably reproduces data in the regions I, II and III. Compared to the predictions of the phenomenological formula of Ref.~\cite{Porto84}, our approach seems show an overall better description of the behaviour of particularly asymmetric systems, such as ${^{16}\textrm{O}}+{^{63}\textrm{Cu}}$. In general, the results of our calculations are in reasonable agreement with the surface friction model calculations of Ref.~\cite{Frobrich84}, while they show a poorer agreement with the predictions of the proximity model Ref.~\cite{Birkelund79}, especially in the region III of the excitation function. Some interesting observations can be made on the position of the maximum of the fusion cross section, $\sigma_f^{max}$. As pointed out in Ref.~\cite{Fonte80}, this position could be important to find the onset of incomplete fusion phenomena. Since this process could be linked on how clustered are the projectile or target, a good reproduction of the position of $\sigma_f^{max}$ is a very important achievement for a fusion model. Our model reproduces fairly well the position of maxima in the data, and is in close agreement with the predictions of the surface friction model on the position in energy of $\sigma_f^{max}$ and its absolute value. Another interesting point is related to the variances of the fusion excitation functions. For nearly-symmetric systems, the variances obtained with our model are a bit smaller than the ones that can be derived from the models of Refs.~\cite{Porto84,Frobrich84}, and similar to the ones obtained with the proximity model of Ref.~\cite{Birkelund79}.

\section{Conclusions}
\label{sec:conclusioni}
In this work, we discuss a novel phenomenological formula for the description of heavy ion fusion cross section data for light to medium mass nuclei, valid in the energy regime ranging from the Coulomb barrier up to the onset of multi-fragmentation processes. The newly proposed model is derived by using state-of-the-art artificial intelligence techniques based on a hybridization of genetic programming and artificial neural networks. The underlying approach is inspired by natural selection in biological systems and allows to perform a global search for the optimal mathematical expression to describe the data even in the presence of a large number of variables and in problems particularly complex to model.

To derive the proposed model, we use an extensive dataset extracted from the NRV database and containing data from more than one hundred collision systems, ranging from light to medium mass systems, subdivided in learning and testing sets. The dataset used to test the model suitably contained a set of collision systems unrepresented in the learning set, thus allowing to probe the generalization capabilities of the new models in the region of light to medium mass systems. In addition, data from heavier systems are also used to test extrapolations of our model outside of the mass range exploited for its derivation. The dataset contained $25$ features, including the inverse of the energy in the center-of-mass frame and several variables used to identify the collision system, as well as nuclear structure properties of projectile and target. The outstanding feature selection capabilities of our approach allowed to suggest that only $3$ of the selected variables can suitably contain the informative content of the whole set of proposed features, and are thus sufficient to predict the complete fusion cross section.

We propose two different models, with different trade-offs between complexity and accuracy. Even if they are particularly computationally simple, both the derived models are capable to give a reasonable description of the whole database in the regions I, II and III. The simplest model has $10$ algebraic nodes and foresees exclusively $3$ features, namely $1/E_{cm}$ and $Z_1Z_2$. The most performing formula ($20$ nodes) ensures the fulfillment of physics boundary conditions for the fusion cross sections at very low (under barrier) and high energies, even if for the heaviest compound systems it is not able to reproduce the trend of sub-barrier data. The results are in reasonable agreement also with phenomenological formulas previously reported in the literature and also with detailed theoretical calculations based on several approaches commonly used in heavy ion fusion: surface friction, proximity, critical distance and compound nucleus limitation models. Our novel investigation suggests that the variables $\frac{1}{E_{cm}}$, $Z_1$, $Z_2$ contain essentially the largest part of the information, for the purpose of predicting $\sigma_{f}$, of all features considered in the analysis. Most likely, the fragmentation of the dataset does not allow to derive further clear dependencies on some of the other structure parameters here adopted as features. However, the present analysis cannot exclude that other nuclear structure characteristics, uncorrelated to those used here to derive the new models, might help to improve the prediction of $\sigma_f$. All these findings undoubtedly demand for new experimental campaigns to improve the existing dataset of heavy-ion fusion cross section from the Coulomb barrier to the multi-fragmentation regime.

In turn, the artificial intelligence approach adopted in this study allowed to obtain a quite manageable formula that is able to give reasonable estimates of the heavy ion complete fusion cross sections for a large variety of collision systems ($A_{tot}\simeq 20-80$) and a broad bombarding energy regime. The new phenomenological model could be particularly useful in the nuclear physics community, for example, in the planning of experiments and in making estimates of non-direct contributions, due to fusion, in multi-nucleon transfer phenomena.

\section*{Acknowledgments}
\label{sec:acknowledgments}
D.D. acknowledges the University of Sassari "Fondo di ateneo per la ricerca 2020". D.D. wishes to acknowledge funding support from the Italian Ministry of Education, University and Research (MIUR) through the "PON Ricerca e Innovazione 2014-2020, Azione I.2 A.I.M., D.D. 407/2018". I.L. acknowledges the INFN SyLiNuRe grant.

%bibliography
\section*{References}
\bibliographystyle{iopart-num}
\bibliography{202110_FusionPaper.bib}

\end{document}